\documentclass[runningheads]{llncs}
\usepackage[T1]{fontenc}
\usepackage{graphicx}
\usepackage{float}
\usepackage{caption}
\usepackage{subcaption}
\usepackage{booktabs}
\usepackage{pifont}
\usepackage{listings}
\usepackage{todonotes}
\usepackage{xurl}
\usepackage{array}
\usepackage{makecell}
\newcommand{\cmark}{\ding{51}}
\newcommand{\xmark}{\ding{55}}
\begin{document}

\title{Hidden Dependencies and Component Variants in SBOM-Based Software Composition Analysis}
\titlerunning{Hidden Dependencies...} 
\author{Shawn Rasheed\and Max McPhee\and Lisa Patterson\and Stephen MacDonell \and Jens Dietrich}

\institute{
Victoria University of Wellington\\
\email{\{shawn.rasheed, max.mcphee, lisa.patterson, stephen.macdonell, jens.dietrich\}@vuw.ac.nz}
}\authorrunning{Rasheed et al.}
\maketitle
\begin{abstract}
Software Bills of Material (SBOMs) have emerged as an important
technology for vulnerability management amid rising supply-chain attacks. 
They represent component relationships within a software product and support software 
composition analysis (SCA) by linking components to known vulnerabilities.
However, the effectiveness of SBOM-based analysis depends on how accurately SBOMs represent component identities and  
actual dependencies in software. This paper studies two mismatch patterns:
hidden code-level dependencies that are not represented as component-level 
dependencies, and component variants (clones) that cannot be identified consistently
by scanners. We show that these mismatches can lead to 
inconsistent vulnerability reporting and inconsistent handling of VEX
statements across popular SBOM-based vulnerability scanners. These results highlight limitations in current SBOM production 
and consumption and motivate richer dependency representation and component identity.
\end{abstract}

\section{Introduction}

Software dependencies between components are inherent to  modern
software. They allow developers to reuse existing code instead of
implementing functionality from scratch, which can improve efficiency and
scalability. However, using third-party components can introduce
risks~\cite{kula2018developers,cox2019,zimmermann2019smallworld,decan2018impact,ohm2020backstabber,prana2021outofsight,alfadel2023python}. For instance, upstream components can contain
security vulnerabilities that propagate and impact software products that
depend on them.

Software Bills of Materials (SBOMs)~\cite{xia2023empirical,stalnaker2024bomsaway,bi2024ontheway}
 attempt to address the transparency
issue in dependency use. An SBOM is a standard document that lists the
ingredients in a software product. 
SBOMs have gained significant traction following the \textit{Log4Shell} vulnerability in Apache Log4j, which exposed how difficult it is to identify affected systems without precise knowledge of software dependencies.  In response, governments have increasingly formalized SBOM requirements, most prominently through U.S. Executive Order 14028~\cite{biden2021eo14028}, which mandates SBOM provision for software supplied to federal agencies.

SBOMs can also report dependency
relationships between components that constitute a product. A vulnerability in a component can propagate across dependencies. However, not all reported
vulnerabilities are exploitable in the context of a software product and
vendors may issue Vulnerability Exploitability eXchange (VEX) statements
to communicate such information~\cite{cisa2022vexusecases,cisa2023vexmin,oasis2022csaf,cyclonedx2025vex,openvex2023spec}. Security tools downstream can consume
VEX information to reduce noise and help prioritise vulnerability
reports.

SBOMs are both produced and consumed by software composition analysis (SCA) tools that scan for security vulnerabilities. Therefore it is important to understand their effectiveness. This paper empirically investigates two mismatch patterns in SBOM-based software 
composition analysis: hidden code-level dependencies and hidden component variants.

First, software may contain hidden code-level 
dependencies that are not represented as component-level dependencies
in manifest-derived SBOMs. Prior work has shown that declared component dependencies are often unused at the code level, creating bloated dependencies~\cite{sotovalero2021bloated,latendresse2022notall,jafari2021dependency,turcotte2025bloated}. This paper studies the reverse pattern: code-level dependencies that have no corresponding component-level declaration in the SBOM. In practice, this pattern arises through transitive or ``deep'' access, where application code directly uses functionality from a transitive dependency without declaring that dependency explicitly. Using such transitive dependences is common in Java projects~\cite{jayasuriya2023breaking}. An example of this is in a BitTorrent library for Java, \texttt{bt}\footnote{\url{https://github.com/atomashpolskiy/bt}}. The component \texttt{bt-dht} calls \texttt{com.google.common\-.io.Files::create\-TempDir} in the Google Guava library. Guava is not a declared dependency of the component and it is available transitively via \texttt{bt-core}. Moreover, the resolved version of Guava (and the called method) has a known vulnerability, CVE-2020-8908. This highlights the importance of correctly mapping code-level dependencies to their dependency sources to determine vulnerability exposure and precisely scope  exploitability.

Second, vulnerable functionality may appear
in component variants (clones) whose identity is not represented or
interpreted consistently by scanners. This pattern is prevalent in Java projects~\cite{dietrich2024blindspots}. A notable vulnerable component is  org.json:json:20230227, with 419 vulnerable component variants found and reported~\cite{dietrich2024blindspots}. The authors also report several clones of org.apache.log\-ging.\-log4j:log4j-core:2.14.1 with the critical CVE-2021-44228 (log4shell). In this paper, component variants are components whose identity differs from an original upstream artifact while remaining derived from it, including clones, shaded components, renamed packages, and other repackaged forms.

We study how these two patterns 
affect vulnerability reporting and VEX handling across current tools. Distinguishing these patterns are   especially important when VEX documents declare a vulnerability non-exploitable for one dependency path, creating the possibility that hidden code-level dependencies cause that vulnerability to remain reachable through a path outside the intended suppression scope.

In this study, we investigate:
\begin{itemize}
\item How hidden code-level dependencies affect SBOM and VEX-based
vulnerability analysis in current SCA tools.
\item How component variants affect SBOM-based vulnerability
detection in current SCA tools.
\end{itemize}
 
\section{Related Work}

\subsection{SCA Challenges - Precision}

Incidents including the Equifax data breach and the leftpad package removal demonstrate difficulties assessing bugs present in component dependency graphs  of an application. They only provide indicatative information and Hejderup et al. \cite{hejderup2018ecosystem} propose a fine-grained dependency network, going beyond packages, into call graphs, resulting in a versioned, ecosystem level call graph.

\subsection{SCA Challenges - Recall}

Schott et al. \cite{schott2025bytecode} present a bytecode-centric dependency scanner for Java, named Jaralyzer. This directly analyses a dependency's bytecode, not relying on metadata or source code of the included OSS dependencies. This study claims that Jaralyzer performs more effectively than other popular dependency scanners including Eclipse Steady by detecting more true vulnerabilities and giving fewer false warnings.

Dietrich et al. \cite{dietrich2023security} investigated blind spots in software composition analysis, with a novel approach presented to detect vulnerabilities in cloned or shaded components. This study demonstrates the revelation of blind spots in vulnerability databases and tools that rely on those. Results indicated the need to design software composition analysis tools which can perform deep analyses, not only relying on project meta-data.

 Eclipse Steady is built to analyse Java applications to detect whether they are dependent upon open-source components that have known vulnerabilities; to collect evidence about execution of vulnerable code using a combination of both static and dynamic analysis techniques; and to support developers to mitigate such dependencies.

\subsection{SBOM/VEX Background and Inconsistency}

Modern applications are built on software supply chains that are complex, incorporating many procedural and technical aspects. These include first-party source code, third-party dependencies, source control processes, tools for building and packaging, and services to manage packages \cite{melara2022software}.

Since 2013, usage of outdated and vulnerable software has ranked in the top 10 of Open Worldwide Application Security Project (OWASP) security risks for  web applications. In 2024, at least four top 10 risks were due to dependency-induced threats \cite{rosso2025practical}.

Currently, OWASP \footnote{\url{https://owasp.org/www-project-open-source-software-top-10/}} highlights Known Vulnerabilities, Compromise of Legitimate Package, Unmaintained Software, Outdated Software, Untracked Dependencies, License Risk, Immature Software, Unapproved Change, and Under/over-sized Dependency as among the top 10 Risks for Open Source Software.

Mirakhorli et al. \cite{mirakhorli2024landscape}  catalog tools in the SBOM ecosystem, analyzing and evaluating them. VEX provides a method to enable users to prioritise critical vulnerabilities, thereby reducing unnecessary mitigations \cite{kern2025ensure}.

However, using VEX is not without challenges. Maratos et. al \cite{maratos2025supply} found there is a shortage in cyber professionals who possess expertise in software supply chain security, analysis of SBOMs and interpretation of VEX documentation. Additionally, some organisations may have neither the funding nor the research to implement SBOM and VEX documentation. Churakova et al. 
\cite{churakova2025vexed} present a study analysing current VEX-generation tools applied to containers, to ascertain the consistency of VEX generation tools including Grype, Trivy, DepScan, Scout, OSV, SNyk and Vexy. Reported results indicate low levels of consistency among the tools, and low levels of maturity in the VEX tooling space.

\subsection{Practitioner Perspectives on Tools}

There is regular commentary on VEX-related issues by industry practitioners on the web. These posts are useful to find the types of issues that are currently being highlighted by those who are using SBOM and VEX tools in practice. 

For example, in Aqua Security’s VEX discussion board, practitioners highlight issues including differences in results for VEX between SBOM and image scanning\footnote{\url{https://github.com/aquasecurity/trivy/discussions/10467}}; and a claim that Trivy  does not manage VEX attestation as expected with the option \texttt{—vex oci}\footnote{\url{https://github.com/aquasecurity/trivy/discussions/9833}}. Another contributor claims that Trivy ignores VEX documents and one notes that VEX is good for image, software or container, but queries whether it also works for library. This thread also suggests it should come with tooling to check there is no indirect dependencies usage\footnote{\url{https://github.com/aquasecurity/trivy/discussions/7784}}. However not all comments are negative, some are constructive, including proposal of a feature idea to complement VEX with a simple, built-in reachability check\footnote{\url{https://github.com/aquasecurity/trivy/discussions/9541}}.

In the OpenVEX discussion board, other constructive suggestions are to add risk contextualisation in OpenVEX statements\footnote{\url{https://github.com/orgs/openvex/discussions/23}}, and for greater OpenSSF WG alignment and spec purpose - a value statement about the need that OpenVEX solves in addition to VEX\footnote{\url{https://github.com/orgs/openvex/discussions/19}}.  One contributor proposes `jvex: Java types for Open VEX documents', the contributor aiming to enhance the listings of available tools for authors and consumers who are using Java\footnote{\url{https://github.com/orgs/openvex/discussions/18}}.

Anchore community's discourse group\footnote{\url{https://anchorecommunity.discourse.group/c/grype/6}} contains discussion around the addition of ignore filters for VEX entries. Discussion\footnote{\url{https://github.com/ossf/cve-bin-tool/issues/5694}} around Cve-Bin-Tool queried whether Cve-Bin-Tool version produced different CVE output.

A 2024 OSV-Scanner discussion\footnote{\url{https://github.com/google/osv-scanner/issues/1031}} queried Java import reachability, while older entries were concerned with automatically generating VEX statements based on call graph analysis or ignored vulnerabilities, plus an issue regarding building an automated/guided remediation feature for OSV-Scanner\footnote{\url{https://github.com/google/osv-scanner/issues/352}}. The OSV developers' blog discusses problems with generating VEX and provides recommendations on ways to implement VEX at scale. There is discussion explaining how OSV can help, including intermediate VEX statements, perhaps via a ``ignore file''-like mechanism \footnote{\url{https://osv.dev/blog/posts/automating-and-scaling-vex-generation/}}.

The OWASP SBOM Forum (VEX subgroup)\footnote{\url{https://owasp.org/www-project-sbom-forum/}} plans to produce VEX specs in both of the formats, moving to develop playbooks describing how to both create and ingest documents in each of the formats, with a possibility of developing prototype tools in each format.

\section{Background}

\subsection{Dependency Models}

For the purposes of this study, we need to distinguish dependencies
among code elements from dependencies among software supply-chain
components. The former are semantic relationships between elements in the
code. For instance in Java, such dependencies are established by field accesses, type references, interface
implementations and method or constructor invocations. 
These
relationships may occur within one component or across multiple
components at the supply-chain level. Supply-chain
components include libraries, packages, and modules that must be
present to build, link, package, deploy, or run the software.
On the other hand, supply-chain level components are declared in component metadata.

\begin{figure}[H]
  \centering
  \includegraphics[width=0.75\textwidth]{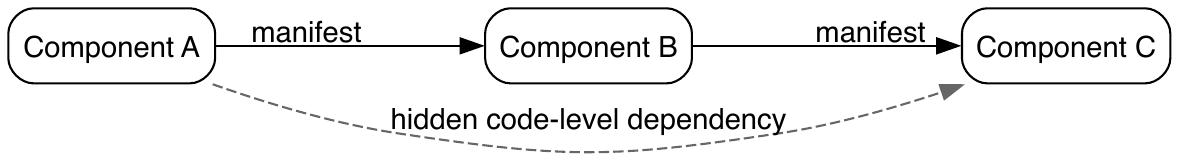}
  \caption{Component-level vs.\ code-level dependency graph.}
  \label{fig:deps}
\end{figure}

Figure~\ref{fig:deps} shows the distinction between component-level and
code-level dependencies. In the component graph (derived from declared
component metadata from manifests such as \texttt{pom.xml},
\texttt{build.gradle} and \texttt{package.json}, or package registries),
A declares a direct dependency on B, and B declares a direct dependency
on C. Thus A depends on C transitively. In the code dependency graph
(only a single edge shown), A has a direct dependency on C. As indicated
earlier, this could be an access to a type in C from A. Even though this
direct code-level dependency is not in A's manifest, it resolves because
C is present during build or runtime, transitively via B.

We identify component-level dependencies declared in manifests as Type-0 dependencies.
We further categorise code-level dependencies as can arise through different mechanisms: static
or dynamic, which we enumerate as:

\begin{itemize}
  \item Type-0: component-level dependencies
  \item Type-1: statically explicit dependencies
  \item Type-2: dynamically dispatched dependencies
  \item Type-3: reflective dependencies
\end{itemize}

SBOM generation relies on component-level dependencies obtained from project manifest files. Direct and transitive component-level dependencies can be constructed
from project metadata. Cross-package code-level dependencies, hidden from
SBOM tools, can appear at source code level (static) or dynamically. 

The rationale for distinguishing between those types here is the impact this has on vulnerability detection. Component-level dependencies are widely used by SCA tools to cross-reference components with vulnerability databases, and then propagate any found vulnerability downstream. This scales very well, but results in a low-precision analysis with high numbers of false positives~\cite{imtiaz2021comparative,liu2023sca,pashchenko2020vuln4real,ponta2020detection,zhan2021dynamics}.  To address this issue, a new generation of SCA tools has emerged, including \textit{coana/socket}~\footnote{\url{https://socket.dev/}} and \textit{endorlabs}~\footnote{https://www.endorlabs.com/}. Here, analysis is based on code-level analysis, in particular call graph construction~\cite{hejderup2018ecosystem,boldi2021finegrained,mir2023transitivity,hejderup2022prazi,keshani2024frankenstein}. Call graph construction is precise for statically explicit dependencies such as Java's static invocation and static field access, but over-approximates dynamically dispatched dependencies (i.e. virtual method invocations) as it needs to perform devirtualisation to model runtime behaviour, resulting in false positives (usually trading off precision for scalability)~\cite{grove1997callgraph,murphy1998empirical,tip2000scalable,lhotak2007comparing}, i.e. spurious call graph edges and reflective dependencies (including reflection, deserialization, dynamic proxies, native method calls etc) are particularly difficult to model soundly by static analysers~\cite{ernst2003synergy,livshits2015soundiness}, usually leading to false negatives, i.e. missing call graph edges~\cite{sui2020recall,chakraborty2022rootcause,lehmann2023toughcall}.

\subsection{Software Bills of Materials (SBOM)}

Software Bills of Materials  SBOMs are machine readable inventories of software components, which includes their versions and dependencies. There are two major formats: CycloneDX (OWASP) and SPDX (Linux Foundation). SBOMs for software can be generated at various points in the lifecycle: build-time (e.g. the CycloneDX Maven plugin for Java), image-scanning (e.g. Syft) and runtime (e.g. jbom). For build-time SBOMs, they capture declared direct and transitive dependencies along with package identifiers (e.g. PURL). They do not capture dynamic/runtime dependencies, which paths or components in the code are actually executed. Vulnerability scanners use SBOMs as input to match component identifiers (e.g. PURL or CPE) against vulnerabilities records from sources such as NVD or GHSA.

\subsection{Vulnerability Exploitability eXchange (VEX)}

Vulnerability Exploitability eXchange (VEX) specifies a machine-readable assertion about whether a known vulnerability is exploitable in a specific product. The purpose of having VEX is to reduce noise from scanners reporting vulnerabilities that are not reachable or applicable in the product's context. Known formats for VEX are CycloneDX VEX, OpenVEX and CSAF VEX.  Per the OpenVEX specification, a VEX statement has the following elements:

\begin{itemize}
  \item \textbf{products}: the distributable artifact(s) being assessed
        (e.g.\ an application or container image)
  \item \textbf{subcomponents}: the vulnerable library/component within
        that product
  \item \textbf{status}: one of these: \texttt{not\_affected}, \texttt{affected}, \texttt{fixed}, \texttt{under\_investigation}
   \item \textbf{justification}: e.g. \texttt{component\_not\_present} or \texttt{vulnerable\_code\_not\_in\_\-execute\_path}
   \item \textbf{impact statements}
\end{itemize}

VEX statements are generated by upstream vendors, CI/CD pipelines and security teams consuming scanner output. As for VEX consumption, scanners can ingest VEX documents to suppress or annotate findings.

\begin{figure}[H]
  \centering
  \begin{subfigure}[b]{0.48\textwidth}
    \centering
    \includegraphics[width=\textwidth]{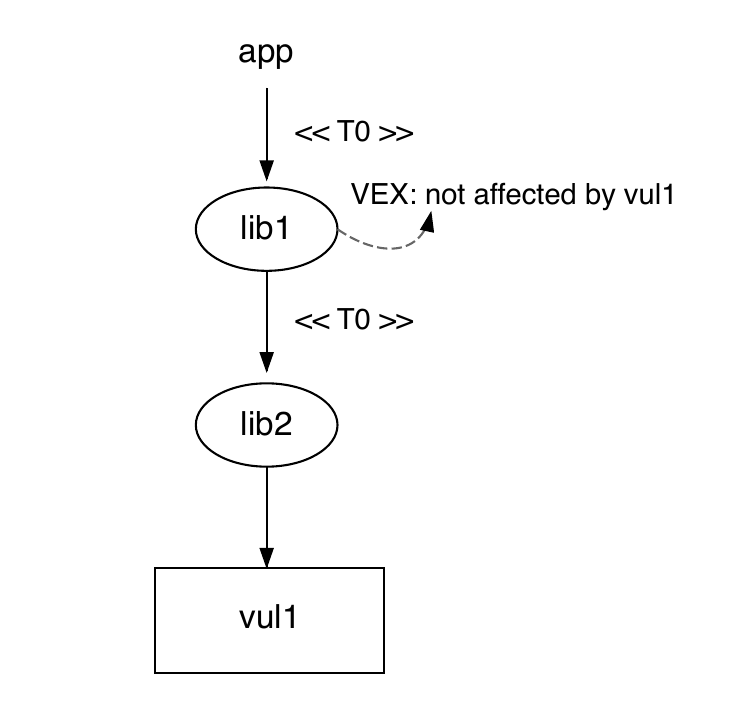}
    \caption{Unreachable vulnerable dependency with VEX statement}
    \label{fig:vex1}
  \end{subfigure}
  \hfill
  \begin{subfigure}[b]{0.48\textwidth}
    \centering
    \includegraphics[width=\textwidth]{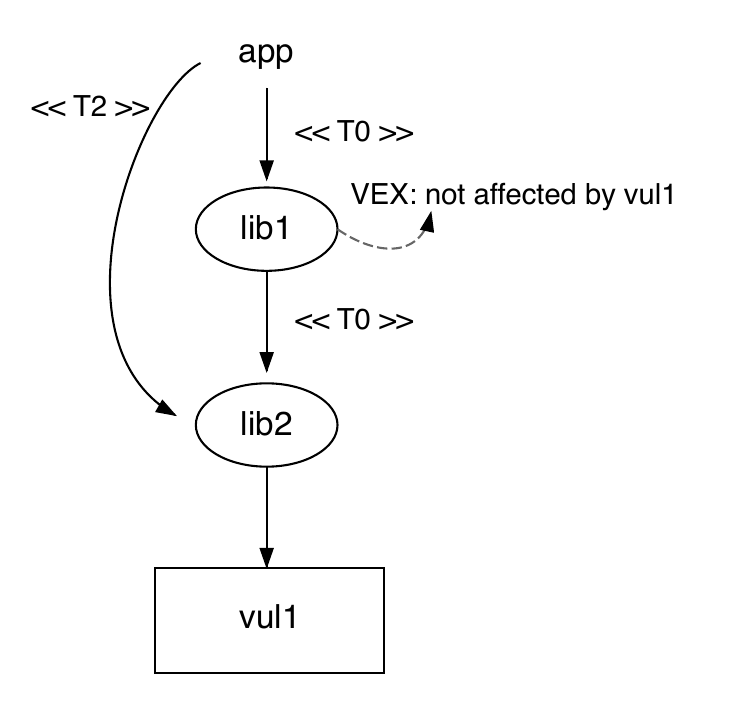}
    \caption{Reachable vulnerable dependency with VEX statement}
    \label{fig:vex2}
  \end{subfigure}
  \caption{VEX suppression scope across dependency paths}
  \label{fig:vex}
\end{figure}

A VEX scope determines the product-component relationship that a VEX statement applies to. The following figures illustrate the scope of a VEX statement for a dependency graph with three components, \texttt{app}, \texttt{lib1} and \texttt{lib2}.
Figure~\ref{fig:vex1} shows a transitive dependency chain for a program, \texttt{app}  depends on \texttt{lib1}  (via a Type-0 build dependency), which depends on  \texttt{lib2}  (Type-0), which contains a known vulnerability, \texttt{vul1}. The dashed arrow from  \texttt{lib1}  represents a VEX \texttt{not\_affected}  statement asserting that  \texttt{lib1}'s use of \texttt{lib2}  does not reach the vulnerability, vul1. This illustrates the basic VEX statement structure:  \texttt{lib1} is the product,  \texttt{vul1}  is the subcomponent, and the status is  \texttt{not\_affected}. Suppression scope: a VEX statement scoped to  \texttt{lib1}  suppresses findings only along the path  \texttt{lib1} $\rightarrow$ \texttt{lib2} $\rightarrow$ \texttt{vul1}. It does not cover other consumers of \texttt{lib2}.

  Figure~\ref{fig:vex2} shows the same chain as Figure~\ref{fig:vex1}, but  \texttt{app}  now also holds a direct Type-2 dependency on \texttt{lib2}. The VEX assertion on \texttt{lib1} remains valid for the dependency on the Type-0 path, but the Type-2 dependency path from  \texttt{app}  to  \texttt{lib2}  is outside its scope. According to the specification\cite{openvex_2025_specopenvexspecmd}, a scanner must still report  \texttt{vul1} as the VEX statement only applies to the product, which is \texttt{lib2}. The specification defines a statement as an assertion about the impact a vulnerability has on a product. This illustrates the suppression scope issue: a VEX statement
  does not transitively suppress all paths to a subcomponent, only the one rooted at its declared product. Whether scanners consistently perform this suppression is the focus of Study I in this paper.

\subsection{Component Variants and Clones}

Cloned, shaded, or otherwise repackaged components introduce a different mismatch problem from hidden code-level dependencies~\cite{dann2022achilles,dietrich2024blindspots,schott2026unshade}. In these cases, the vulnerable code is present in the software, but the component carrying that code no longer has the original package identity expected by vulnerability databases and scanners. This can happen when a library is copied, renamed, relocated, or modified before redistribution. 
This is a widely used practice in Software Engineering, with legitimate use cases like shading in Java in order to avoid classpath conflicts. The current practice by AI code generators to avoid dependencies and instead inline code is likely to increase the prevalence of such practices.

\begin{figure}[H]
  \centering
  \includegraphics[width=0.48\textwidth]{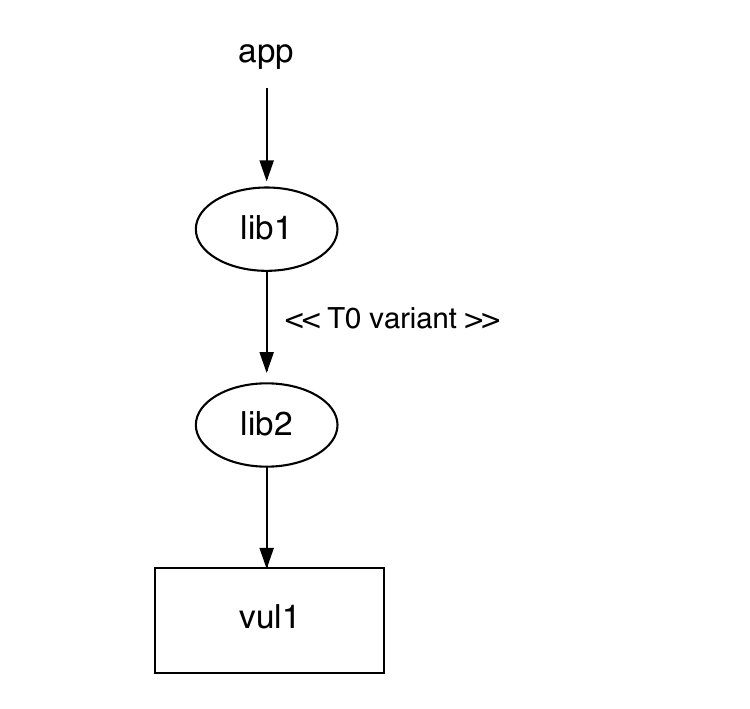}
  \caption{Type-0 component variant: a repackaged component sharing code with an upstream artifact but carrying a different package identity.}
  \label{fig:type0variant}
\end{figure}

Figure~\ref{fig:type0variant} depicts this relationship in the dependency graph as a Type-0 edge.
As a result, with the loss of the original identity information, matching based only on standard package identifiers such as PURLs or CPEs may fail, even though the cloned component remains functionally derived from a known vulnerable upstream artifact. SBOM formats attempt to address this problem by recording lineage or variant relationships, for example through \texttt{pedigree.variants} in CycloneDX or hasVariant relationships in SPDX. Whether scanners consume this information consistently remains an open practical question and is the focus of Study II in this paper.

\section{Study I}

\subsection{Design}

A benchmark is designed to isolate cases where the application has a direct code-level dependency on a vulnerable transitive component, even though that dependency is absent from the component-level SBOM.

The benchmark consists of four application variants designed to exercise Type-1, Type-2 and Type-3 hidden dependencies. Each application depends on a library, \texttt{lib1}, which in turn depends on \texttt{log4j-core:2.8.1}, which has the \textit{Log4Shell} vulnerability. This mirrors the component-dependency graph in Figure~\ref{fig:vex1} for the test case with no code-level dependency outside component-dependencies and Figure~\ref{fig:vex2} for Types-1 to 3. The project files are available on a GitHub repository\footnote{\url{https://github.com/ssc-fort/sbom-vex-benchmarks}}

\subsubsection{VEX statement for library, \texttt{lib1}}

declares that  \texttt{lib1}  is  \texttt{not\_affected} by the vulnerability in its dependency on \texttt{log4j-core:2.8.1}.

\subsubsection{Test cases}
\paragraph
\noindent\resizebox{\linewidth}{!}{\begin{tabular}{|l|l|l|l|l|}
\hline
\textbf{Test case} & \textbf{Dependency Pattern} & \textbf{Vulnerable?}  & \textbf{Expected scanner result} \\
\hline
app-static & Type 1 & Yes (explicit type reference)  & detected \\
\hline
app-dynamic & Type 2 & Yes (interface only at call site)  & detected \\
\hline
app-reflective & Type 3 string literal & Yes (string only)  & detected \\
\hline
app-unreachable & No hidden code-level dependencies & No call path to vulnerable class  & not-detected \\
\hline
\end{tabular}}
\captionof{table}{Test cases and expected detection behavior}
\label{tab:test-cases}
\vspace{1cm}
The benchmark includes a script to score each SCA tool's detected/not-detected result against the expected result.

\subsubsection{SBOM Format Selection}

For the SBOMs, this study uses CycloneDX as the format because component-level relationship types are not relevant here; that aspect is examined in Study II. For this section of the experiment, consistency of accepted formats and preservation of component-level dependency context are sufficient. In addition, CycloneDX is supported by a Maven plugin, which makes it suitable for this study.

\subsection{SCA Tool Selection}

We select SCA tools that support SBOM-based vulnerability detection and which can also consume VEX information. This narrows down the list of tools to three: Grype, Trivy and CVE-Bin-Tool. These tools are considered state-of-the-art and used in several studies \cite{odonoghue2024saner,churakova2025vexed}. This helps us answer the first question, which addresses how SCA tools handle SBOM and VEX-based detection. It is important to note that the tools do not consider code-level dependencies and the expected behaviour is for the tools to comply with the specification with regard to applying VEX in filtering vulnerability reports. Note that Table~\ref{tab:test-cases} shows the desired behavior of tools that take code-level dependencies into account when scanning for vulnerabilities.

\subsection{Tool Configurations}

Each scanner was invoked against the CycloneDX SBOM (\texttt{target/bom.json}) generated for each
test application. The tool versions and configurations are listed in Table~\ref{tab:sca-tools-phases-config}. Scans were run in two phases: once without any VEX input (base detection) and
once with \texttt{lib1.vex.json} supplied as a VEX document (suppression phase). Grype performs content-based format detection. OpenVEX documents are passed via the \texttt{--vex}
flag. Trivy uses file-extension-based format detection. OpenVEX documents are passed via the
\texttt{--vex} flag. CVE-Bin-Tool auto-detects CycloneDX SBOMs when passed directly. OpenVEX documents are supplied
via \texttt{--vex-file} with \texttt{--vex-type openvex}; the \texttt{--filter-triage} flag
instructs the tool to suppress findings covered by the VEX triage. The OSV data source was
disabled to avoid network dependency on the OSV dataset during scanning.

\begin{table}[h]
  \centering
  \small
  \begin{tabular}{p{2.5cm} p{1.5cm} p{5.3cm} p{5.8cm}}
    \hline
    \textbf{Tool} & \textbf{Version} & \textbf{Phase 1 configuration} & \textbf{Phase 2 configuration} \\
    \hline
    Grype & 0.109.0 &
    \makecell[l]{\texttt{grype sbom:<app>/target/bom.json} \\ \texttt{-o json}} &
    \makecell[l]{\texttt{grype sbom:<app>/target/bom.json} \\ \texttt{--vex lib1/lib1.vex.json} \\ \texttt{-o json}} \\
    \hline
    Trivy & 0.69.3 &
    \makecell[l]{\texttt{trivy sbom <app>/target/bom.json} \\ \texttt{--format json}} &
    \makecell[l]{\texttt{trivy sbom <app>/target/bom.json} \\ \texttt{--vex lib1/lib1.vex.json} \\ \texttt{--format json}} \\
    \hline
    CVE-Bin-Tool & 3.4 &
    \makecell[l]{\texttt{cve-bin-tool <app>/target/bom.json} \\
                 \texttt{--disable-data-source OSV} \\
                 \texttt{--format json} \\
                 \texttt{--output-file <output>.json}} &
    \makecell[l]{\texttt{cve-bin-tool <app>/target/bom.json} \\
                 \texttt{--vex-file lib1/lib1.vex.json} \\
                 \texttt{--vex-type openvex} \\
                 \texttt{--filter-triage} \\
                 \texttt{--disable-data-source OSV} \\
                 \texttt{--format json} \\
                 \texttt{--output-file <output>.json}} \\
    \hline
  \end{tabular}
  \caption{SCA tools, versions, and scan-phase configurations used in the experiment.}
  \label{tab:sca-tools-phases-config}
\end{table}

\subsection{Results}

\subsubsection{Phase 1: Base Detection}

Table~\ref{tab:study1-base-results} shows whether CVE-2017-5645 was detected by each scanner
without any VEX input. All three scanners flagged the vulnerability in every test application,
including \texttt{app-reachable}, which has no call path to the vulnerable class. This is expected
for dependency-only scanners that report on artifact presence rather than reachability.

\begin{table}[h]
  \centering
  \begin{tabular}{lcccc}
  \toprule
  \textbf{Tool} & \textbf{app-static} & \textbf{app-dynamic} & \textbf{app-reflective} & \textbf{app-unreachable} \\
  \midrule
  Grype        & \cmark & \cmark & \cmark & \cmark \\
  Trivy        & \cmark & \cmark & \cmark & \cmark \\
  CVE-Bin-Tool & \cmark & \cmark & \cmark & \cmark \\
  \midrule
  Expected     & \cmark & \cmark & \cmark & \xmark \\
  \bottomrule
  \end{tabular}
  \caption{Phase 1 base detection of CVE-2017-5645 by scanner and test case.
           \cmark~= detected, \xmark~= not detected.}
  \label{tab:study1-base-results}
\end{table}

\subsubsection{Phase 2: VEX Suppression}

Table~\ref{tab:study1-vex-results} shows results when \texttt{lib1.vex.json} was supplied to each
scanner. The VEX document asserts that \texttt{lib1} (not the downstream applications) is
\texttt{not\_affected} by CVE-2017-5645. A correct scanner should therefore suppress the finding
only for \texttt{app-unreachable} (which never reaches the vulnerable code) and retain it for the
three applications that do.

\begin{table}[h]
  \centering
  \begin{tabular}{lcccc}
  \toprule
  \textbf{Tool} & \textbf{app-static} & \textbf{app-dynamic} & \textbf{app-reflective} & \textbf{app-unreachable} \\
  \midrule
  Grype        & \cmark & \cmark & \cmark & \cmark \\
  Trivy        & \xmark & \xmark & \xmark & \xmark \\
  CVE-Bin-Tool & \cmark & \cmark & \cmark & \cmark \\
  \midrule
  Expected     & \cmark & \cmark & \cmark & \xmark \\
  \bottomrule
  \end{tabular}
  \caption{Phase 2 detection of CVE-2017-5645 with VEX suppression applied.
           \cmark~= detected, \xmark~= not detected.}
  \label{tab:study1-vex-results}
\end{table}

This is the core Pattern 1 failure: suppression is applied according to incomplete component-level information, while the relevant code-level dependency remains outside the SBOM path used by the scanner.

No scanner achieved the correct outcome across all four test cases. Grype and CVE-Bin-Tool matched
VEX on the product PURL (\texttt{pkg:maven/example/lib1@1.0-SNAPSHOT}); since the scanned
artifacts are the downstream applications rather than \texttt{lib1} itself, no PURL match occurs
and suppression is never applied — leaving \texttt{app-unreachable} incorrectly flagged. Trivy
matched VEX on the subcomponent PURL (\texttt{log4j-core@2.8.1}), suppressing the finding
globally across all four applications and incorrectly clearing the three cases where the
vulnerable code is reachable.

\section{Study II}

\subsection{Design}

A Study II benchmark evaluates whether scanners can detect vulnerabilities in a component variant represented in SBOM metadata. In this benchmark, the variant is implemented as a clone of org.apache.logging.log4j/log4j-api@2.10.0

\subsubsection{Test cases}

The benchmark uses SBOMs  derived for a Java application that uses a cloned component \texttt{uk.co.nichesolutions.logging.log4j/log4j-api@2.6.3-CUSTOM}. 
This component is a clone of \texttt{org.apache.logging.log4j/log4j-api@2.10.0}, which has known security vulnerabilities, CVE-2021-44228 \footnote{\url{https://github.com/advisories/GHSA-jfh8-c2jp-5v3q}} and CVE-2021-45046 \footnote{\url{https://github.com/advisories/GHSA-7rjr-3q55-vv33}}. The SBOM files are available in the same repository cited in Study I.

Four SBOM files were generated to assess the ability of tools to detect vulnerabilities using variant metadata:

\begin{itemize}
\item \texttt{sbom-cycloneDX.json} - Generated in CycloneDX format, without any variant metadata
\item \texttt{sbom-cycloneDX-with-variants.json} - Generated in CycloneDX format, including variant metadata stored within the \texttt{components[].pedigree.variants[]} array. This array contains information pertaining to the original package.
\item \texttt{sbom-spdx.json} - Generated in SPDX format, without any variant metadata
\item \texttt{sbom-spdx-with-variants.json} - Generated in SPDX format with variant metadata represented as a \texttt{relationships[]} entry with \texttt{relationshipType: "hasVariant"} linking the cloned package to the original. 
\end{itemize}

\subsection{Tool Versions and Configurations}

For the variants tests, each scanner was invoked against all four SBOM files using versions and configurations listed in Table~\ref{tab:study2-scanner-configurations}. Grype performs content-based format detection and requires no explicit format flag when scanning
an SBOM file. Trivy uses file-extension-based format detection, distinguishing CycloneDX (\texttt{.cdx.json})
from SPDX (\texttt{.spdx.json}). Files were renamed accordingly prior to scanning. OSV-Scanner uses the \texttt{scan} subcommand and also relies on file-extension-based format detection. Files were renamed with the appropriate extensions prior to scanning. CVE-Bin-Tool requires the SBOM format to be specified explicitly via the \texttt{--sbom} flag. The OSV data source was disabled to isolate results to NVD and other non-OSV sources.

\begin{table}[h]
  \centering
  \small
  \begin{tabular}{p{2.8cm} p{1.5cm} p{9.5cm}}
    \hline
    \textbf{Tool} & \textbf{Version} & \textbf{Configuration} \\
    \hline
    Grype & 0.109.0 &
    \makecell[l]{\texttt{grype sbom:<sbom-file>.json} \\ \texttt{-o json}} \\
    \hline
    Trivy & 0.69.3 &
    \makecell[l]{\texttt{trivy sbom --cache-dir /cache/trivy} \\
                 \texttt{--format json <sbom-file>.<ext>.json}} \\
    \hline
    OSV-Scanner & 2.3.3 &
    \makecell[l]{\texttt{osv-scanner scan --sbom} \\
                 \texttt{<sbom-file>.<ext>.json --format json}} \\
    \hline
    CVE-Bin-Tool & 3.4 &
    \makecell[l]{\texttt{cve-bin-tool --sbom <cyclonedx|spdx>} \\
                 \texttt{--sbom-file <sbom-file>.json} \\
                 \texttt{--disable-data-source OSV} \\
                 \texttt{--format json} \\
                 \texttt{--output-file <output-file>.json}} \\
    \hline
  \end{tabular}
  \caption{Security scanners, versions, and configurations used in the experiment.}
  \label{tab:study2-scanner-configurations}
\end{table}

\subsection{Results}

\begin{table}[h]
  \centering
  \begin{tabular}{lcccc}
  \toprule
  \textbf{Tool} & \textbf{CycloneDX} & \textbf{CycloneDX + variants} & \textbf{SPDX} & \textbf{SPDX + variants} \\
  \midrule
  trivy        & \xmark & \xmark & \xmark & \cmark \\
  grype        & \xmark & \xmark & \xmark & \cmark \\
  osv-scanner  & \xmark & \xmark & \xmark & \cmark \\
  cve-bin-tool & \xmark & \xmark & \xmark & \xmark \\
  \bottomrule
  \end{tabular}
  \caption{Detection of CVE-2021-44228 or CVE-2021-45046 by scanner and SBOM variant.
           \cmark~= detected, \xmark~= not detected.}
  \label{tab:variant-results}
  \end{table}

 The vulnerable component is present, but scanner behavior depends on whether variant lineage is represented and consumed consistently across SBOM formats.

Three tools detected vulnerabilities only for the SPDX SBOM containing variant metadata. This suggests partial support for variant-aware matching in SPDX-based inputs. In contrast, none of the evaluated tools detected the cloned component in either CycloneDX input, indicating that \texttt{pedigree.variants} was not used effectively in this benchmark. 
\section{Discussion}

\subsection{Hidden code-level dependencies and VEX scope}

Study I shows that hidden code-level dependencies can break the intended scope of VEX suppression when scanners rely only on component-level matching. In the benchmark, all tools reported the vulnerable component before VEX suppression, including the unreachable case, because the vulnerable artifact was present transitively in the SBOM. After VEX was supplied, no evaluated scanner applied suppression in a way that matched the expected path-sensitive interpretation. Grype and CVE-Bin-Tool failed to suppress even the unreachable case, while Trivy suppressed all cases, including those where vulnerable code remained reachable. This suggests that current VEX handling in SBOM-based scanning does not consistently respect the distinction between product scope and dependency-path scope.

\texttt{app-unreachable} shows that any tool doing component-level scanning reports the vulnerability. In large projects, this effect multiplies with a larger number of transitive dependencies. Scanners may refine reports using reachability analysis. VEX is a form of manual mitigation to reduce false positive noise. However, the benchmark shows that this can be prone to introducing false negatives.

\subsection{Cloned/component variants and identity matching}

Unlike Study I, this second pattern does not arise from a missing edge in the dependency graph, but from inconsistent handling of component lineage and identity at the SBOM level.

Study II shows a different way in which SCA can fail on accuracy: the vulnerable code may be present, but under a component identity that scanners do not match reliably against vulnerability records. In the benchmark, three tools reported the vulnerability only for the SPDX SBOM containing variant metadata, while none detected the cloned component in either CycloneDX case. This indicates that variant-aware matching remains format and tool dependent. The result should  be interpreted as evidence of inconsistent support for component lineage, rather than conclusive proof of correct \texttt{hasVariant} semantics.

\subsection{Implications for SBOM producers and scanners}

Together, the two studies show that SBOM-based SCA fails when the dependency model used by scanners diverges from actual dependencies and component identity in software. They also show that current SBOM production and consumption pipelines are limited by the granularity of the dependency and identity information they exchange. For hidden code-level dependencies, manifest-derived SBOMs omit edges that may matter for reachability and exploitability assessment. For cloned or component variants, standard identifiers alone may be insufficient to preserve lineage to vulnerable upstream artifacts. Scanner implementations would also benefit from clearer rules for handling VEX scope and for consuming variant metadata consistently across formats.

\subsubsection*{Proof of concept: bytecode-level SBOM enrichment}

To explore the feasibility of addressing the hidden code-level dependency
pattern, we implemented a proof-of-concept tool 
that enhances an existing CycloneDX SBOM with dependency edges derived from
bytecode analysis. The tool accepts a SBOM and
a set of compiled JARs, and builds a mapping from JVM internal class names to
SBOM component identifiers using each JAR's embedded Maven coordinates, and
then scans each class file for named type references: method call owners,
field owners, \texttt{new} expressions, casts, and class literals. Any
cross-component reference that is absent from the SBOM's \texttt{<dependencies>}
section is written back as an enrichment edge, annotated with the call site and
the analysis technique used.

Applied to the Study~I benchmark, the tool correctly identifies the missing
\texttt{app-static} $\rightarrow$ \texttt{log4j-core:2.8.1} edge that the
CycloneDX Maven plugin does not produce, and it produces no spurious edges for
\texttt{app-unreachable}. This static reference scan is, by design, limited to type
names that appear in the bytecode constant pool: it covers the Type-1
case (\texttt{app-static}) but not the Type-2 virtual-dispatch case
(\texttt{app-dynamic}) or the Type-3 reflection case (\texttt{app-reflective}).
Advanced program analysis will be required to resolve those patterns precisely.

This suggests that enriching SBOMs at the bytecode level is both
tractable and effective for the statically resolvable subset of hidden
dependencies. An enriched SBOM would give downstream scanners the missing
edge through which to evaluate VEX scope, potentially enabling the
path-sensitive suppression behaviour that none of the evaluated tools exhibited.
The degree to which full call-graph enrichment, covering virtual dispatch and
reflection, can be made practical within build pipelines and how 
scanner and VEX implementations would consume such edges correctly, remains a question for future research.

\subsubsection{Generalisation to Other Ecosystems}

The two mismatch patterns examined in this study are not specific to Java and its ecosystem. The use of transitive dependencies extends to other languages and ecosystems. For instance, \textit{endorlabs} discussess ``phantom dependencies'' as a problem in Python\footnote{\url{https://www.endorlabs.com/learn/dependency-resolution-in-python-beware-the-phantom-dependency}} and JavaScript\footnote{\url{https://www.endorlabs.com/learn/javascript-typescript-nodejs-reachability-phantom-dependency-detection}}. One manifestation of the ``phantom dependency'' problem they describe is the use of transitive dependencies. Mechanisms that can introduce variants are a known issue in software engineering. These include code cloning through practices such as copying code from websites and using AI tools for code generation \cite{10.1145/3729397}, vendoring, which is closely related to shading, is a practice in Python, Go and PHP.

\subsection{Limitations}

Our studies are limited to Java projects, but the features and tools we evaluated are language-agnostic as they focus on SBOM and VEX handling. Similar behavior is expected for SBOMs from other languages/ecosystems. We use a single vulnerability in the test cases but this does not affect our measurements as we use a vulnerability that is in the vulnerability databases used by all the SCA tools evaluated. Moreover, the goal of the measurement is not accuracy, but the consistency of SBOM and VEX feature handling across tools.

\section{Conclusion}

This paper examined two mismatch patterns in SBOM-based software composition analysis: hidden code-level dependencies and cloned or component variants. The first pattern affects how vulnerabilities are reported and how VEX suppression is applied when code-level reachability is not reflected in component-level dependency graphs. The second affects vulnerability matching when vulnerable code appears under a different component identity from the one represented in vulnerability databases. Across the evaluated tools, neither pattern was handled consistently. These results suggest that  SBOM production and consumption pipelines require richer support for dependency structure, suppression scope, and component lineage in order to support more accurate vulnerability analysis. 
\bibliographystyle{splncs04}
\bibliography{references}
\end{document}